\documentclass{ieeeaccess}
\usepackage{cite}
\usepackage{amsmath,amssymb,amsfonts}
\usepackage{algorithmic}
\usepackage{graphicx}
\usepackage{textcomp}

\usepackage{graphicx}
\usepackage{gensymb}
\usepackage{xurl}
\usepackage{tabu}
\usepackage{multirow}

\newcommand{\gomez}{G\'{o}mez}

\newcommand{\todo}[1]{\iffalse #1 \fi}

\def\BibTeX{{\rm B\kern-.05em{\sc i\kern-.025em b}\kern-.08em
    T\kern-.1667em\lower.7ex\hbox{E}\kern-.125emX}}
\begin{document}
\todo{update before journal submission}
\doi{N/A}

\title{PAseos Simulates the Environment for Operating multiple Spacecraft}
\author{\uppercase{Pablo G\'{o}mez}\authorrefmark{1,}\authorrefmark{2},
\uppercase{Johan \"{O}stman}\authorrefmark{2}, 
\uppercase{Vinutha Magal Shreenath}\authorrefmark{2},
\uppercase{and Gabriele Meoni}\authorrefmark{2,}\authorrefmark{3}} 
\address[1]{Advanced Concepts Team, European Space Agency, European Space Research and Technology Centre (ESTEC), Keplerlaan 1, 2201 AZ Noordwijk, The Netherlands (e-mail: pablo.gomez@esa.int)}
\address[2]{AI Sweden, Lindholmspiren 11, 417 56 Göteborg, Sweden (e-mail: johan.ostman@ai.se, vinutha@ai.se)}
\address[3]{$\Phi$-Lab, European Space Agency, ESA Centre for Earth Observation (ESRIN), Via Galileo Galilei 1, Frascati (00044), RM, Italy (e-mail: gabriele.meoni@esa.int)}
\tfootnote{The work of Johan \"Ostman was funded by Vinnova under grant 2020-04825 and the work of Vinutha Magal Shreenath under Vinnova grant 2021-03643.}

\markboth
{\gomez{} \headeretal: PAseos Simulates the Environment for Operating multiple Spacecraft}
{\gomez{} \headeretal: PAseos Simulates the Environment for Operating multiple Spacecraft}

\corresp{Corresponding author: Pablo \gomez{} (e-mail: pablo.gomez@esa.int).}

\begin{abstract}
The next generation of spacecraft is anticipated to enable various new applications involving onboard processing, machine learning and decentralised operational scenarios.
Even though many of these have been previously proposed and evaluated, the operational constraints of real mission scenarios are often either not considered or only rudimentary.

Here, we present an open-source Python module called PASEOS that is capable of modelling operational scenarios involving one or multiple spacecraft. 
It considers several physical phenomena including thermal, power, bandwidth and communications constraints as well as the impact of radiation on spacecraft. 
PASEOS can be run both as a high-performance-oriented numerical simulation and/or in a real-time mode directly on edge hardware. 
We demonstrate these capabilities in three scenarios, one in real-time simulation on a Unibap iX-10 100 satellite processor, another in a simulation modelling an entire constellation performing tasks over several hours and one training a machine learning model in a decentralised setting. 
While we demonstrate tasks in Earth orbit, PASEOS is conceptually designed to allow deep space scenarios as well.

Our results show that PASEOS can model the described scenarios efficiently and thus provide insight into operational considerations. 
We show this in terms of runtime and overhead as well as by investigating the modelled temperature, battery status and communication windows of a constellation. 
By running PASEOS on an actual satellite processor, we showcase how PASEOS can be directly included in hardware demonstrators for future missions. 

Overall, we provide the first solution to holistically model the physical constraints spacecraft encounter in Earth orbit and beyond. 
The PASEOS module is available open-source online together with an extensive documentation to enable researchers to quickly incorporate it in their studies.
\end{abstract}


\titlepgskip=-21pt

\maketitle

\section{Introduction}
\label{sec:introduction}
\PARstart{T}{he} last two decades have been characterized by significant changes in the availability, types, and usage of spacecraft, especially satellites in Earth orbit and beyond \cite{lemmens2022esa,goldberg2019juventas}. Miniaturization has enabled operating satellites on the scale of centimeters with just a few kilograms of mass \cite{liddle2020space}. 
At the same time, the cost of launching satellites has plummeted, which opens up various new applications for both private and public actors \cite{jones2018recent}.
Furthermore, with the advent of constellations for Earth Observation and with the increased availability of computation, tools to deal with, e.g.,  scale, complexity, and coordination of such distributed systems in space are required. 

Many of the anticipated future applications are related to data processing and transmission \cite{lemmens2022esa,kodheli2020satellite}. 
Indeed, with the rise of artificial intelligence methods in virtually all domains, the next generation of satellites is likely to be equipped with novel tools and hardware \cite{bruhn2020enabling} to enable machine learning methods and other technological innovations such as inter-satellite links (ISL) \cite{giorgi2019advanced}. To this end, a growing corpus of research has been exploring novel applications suitable for this new paradigm, ranging from machine learning applications in remote sensing \cite{gomez2021msmatch,izzo2022geodesy} to federated learning in low-Earth orbit (LEO) constellations \cite{razmi2022board,matthiesen2022federated}. 
However, there are fundamental constraints in the operations of LEO satellites related to, e.g., short communication windows with ground stations (resulting in large latencies and poor availability) \cite{furano2020towards}, temperature, radiation, and power budgets~\cite{wertz1999space}. In addition to a hostile environment, other challenges include increased hardware complexity, and absence of a communication network \cite{curzi2020large}.
Similar constraints exist beyond LEO and, thus, for robust and realistic scenario planning, it is necessary to properly consider them already both in algorithmic design and the demonstration of space applications. Specialized simulation tools exist for many aspects of onboard operation, such as communications \cite{varga2010overview,riley2010ns} or orbital dynamics \cite{maisonobe2010orekit}, but there is a lack of holistic simulation of the environment.

Here, we present an open-source software tool called PASEOS that is dedicated to realistic modelling of various constraints induced by the harsh environment of space. 
PASEOS is based on a modular design and is compatible with existing code in both discrete-event and time-based simulations. 
It supports a variety of scenarios ranging from modelling individual satellites to large-scale constellations. 
It can be run like a classical numerical simulation minimizing time-to-solution or in a real-time mode based on a clock, e.g., the system clock the hardware is running on. 
PASEOS is implemented in Python and is completely agnostic to any machine learning frameworks, such as PyTorch \cite{NEURIPS2019_9015} or TensorFlow \cite{tensorflow2015-whitepaper}, or even the specific application. 
In this work, we describe the system design and mathematical modelling inside of PASEOS. Further, we demonstrate its capabilities on three applications. 
First, by using actual satellite hardware (Unibap iX-10 100 satellite processor), we demonstrate real-time modelling of onboard detection of volcanic eruptions using a single satellite.
Second, we model the operational behavior and constraints of a large LEO constellation limited by thermal, power and communication constraints. 
In the final example, we demonstrate a fully decentralized machine learning application on two satellites modelled with PASEOS using heterogeneous data. 

PASEOS source code is open-source available online.\footnote{\url{https://github.com/aidotse/PASEOS}} We hope to inspire and enable a broad variety of follow-up research by making PASEOS modular, user-friendly and efficient.
Thus, the contributions of this work are:
\begin{itemize}
    \item Creation of the open-source Python framework PASEOS to model the impact of constraints imposed by the space environment
    \item Demonstration that the framework can be used in a wide range of scenarios including operations on actual satellite hardware for real-time implementations, for modelling constellations, and to study distributed machine learning approaches
    \item Detailed analysis of the runtime overhead of PASEOS models in a scenario studying the detection of volcano eruptions on Unibap iX-10 100 processors
    \item Demonstration of the direct impact operational constraints can have on machine learning methods due to power budgets and communication windows
\end{itemize}

\section{Methods}

This section describes the physical models implemented in PASEOS as well as some of the design considerations. In the following, any type of device, vehicle or similar, modelled in PASEOS, will be referred to as an \textit{actor}.

\subsection{Modelling of Spacecraft Constraints}\label{sec:constraints}
Operating spacecraft poses vastly different challenges than operating any type of vehicle or asset on Earth \cite{lentaris2018high, furano2020towards, fortescue2011spacecraft}. 
Here, we briefly outline the constraints considered by PASEOS and how they are modelled. Overall, PASEOS aims to strike a compromise between computational complexity and physical fidelity that enables it to be run in parallel to perform an operation, e.g., the training of a neural network, and yet still account for various constraints on different hardware devices, including embedded systems. 

\begin{figure*}[!htbp]
\centering
\includegraphics[width=\linewidth]{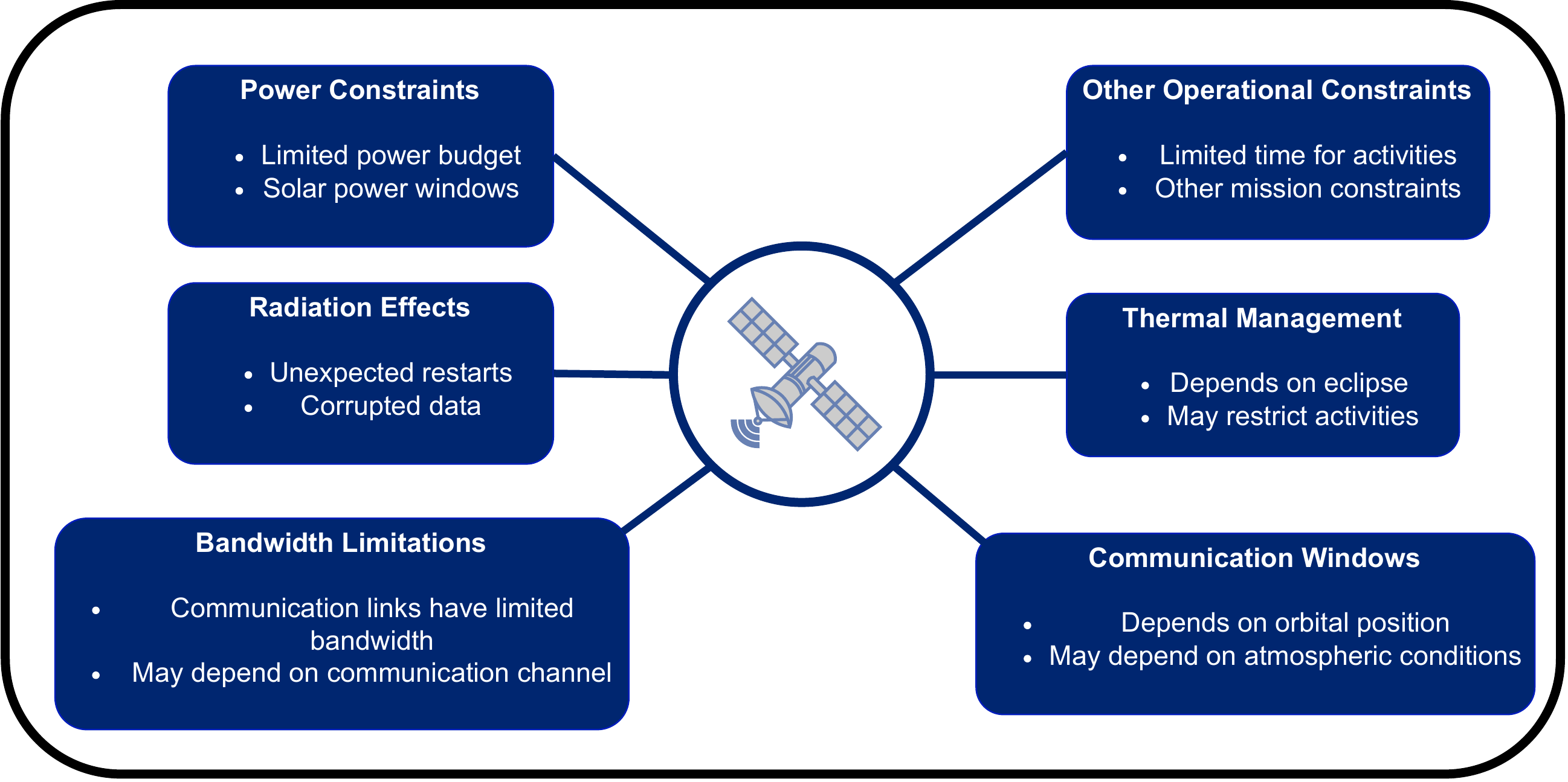}
\caption{Overview of the constraints modelled in PASEOS \label{fig:constraints}}
\end{figure*}

Overall, both physical constraints, such as thermal management or the availability of communication links, as well as operational constraints, such as per-user allocated time slots in missions, can be modelled with PASEOS.  An overview of the modelled constraints can be found in Fig.~\ref{fig:constraints}. In this section, most constraints are explained in relation to orbits around Earth. They translate, however, to other scenario, such as orbits around other celestial bodies or deep space missions.

\subsubsection{Astrodynamics Modelling}
Many of the following constraints depend on the exact ephimerides, i.e., positions and velocities, of the spacecraft and relevant celestial bodies, such as the Sun or the body the spacecraft may be orbiting. In PASEOS, we assume each actor to have a \textit{central body} that it is gravitationally bound to, e.g., the Earth or the Sun. In its current version 0.1.2, PASEOS uses pykep \cite{izzo2012pygmo} to model Keplerian orbits around the central body. The JPL low-precision ephemerides\footnote{\url{https://esa.github.io/pykep/documentation/planets.html#pykep.planet.jpl_lp} Accessed: 2023-01-31} are used to model the position of the Earth and Sun. 
Even though we focus on simple Keplerian orbits in this work, it is conceivable to employ more complex dynamics models ranging from propagators such as SGP4 \cite{vallado2008sgp4} to high-fidelity modelling through \textit{orekit} \cite{maisonobe2010orekit}. 
All positions and velocities in PASEOS are modelled in the inertial frame of the central body.

\subsubsection{Communication Windows}
One of the central constraints in spacecraft operations, especially in LEO, are communication windows during which a spacecraft is able to communicate with ground stations on Earth or with other spacecraft. 
In practice, these windows can depend on a large variety of factors such as the type of communication channel (e.g. optical, radio), environmental factors, such as atmospheric conditions, and others \cite{OpticalCommunication}. 

In PASEOS, we consider communications to be limited by visibility. 
For simplicity and computational efficiency, the version 0.1.2 considers only this constraint for communication between spacecraft. 
Thus, it is computed whether the sphere bounding the central body is obstructing the view between actors. 
For the availability of a ground link between a spacecraft and ground station, we utilize the Skyfield Python module \cite{rhodes2020skyfield} to compute the angle of a spacecraft above the horizon. 
Further, Skyfield is also used to account for ground stations' movement due to the Earth's rotation. 
Ground stations on other celestial bodies are not yet supported. 
Thus, based on these assumptions, PASEOS can be used to compute the availability and duration of communication windows between actors.

\subsubsection{Data Transmission Rate and the Communication Channel}\label{sec:tx_bw}
Aside from the availability of a communication link, a second factor for data transmission is related to the effects of the communication channel, that affect the effective data transmission rate available for the user and impact the quality of data transmitted. 
Especially optical links from satellites in orbit around Earth to ground are highly dependent on atmospheric conditions \cite{kaushal2016optical, OpticalCommunication}. Similarly, space weather events like coronal mass ejections or solar flares, e.g., can impact transmissions to Earth \cite{carrano2009impacts}. These effects can introduce errors in the communication, which could lead to re-transmissions of data packets or require additional data redundancy, which impact the effective downlink data rate available to the user. Additionally, other factors such as the used transmission modulation, channel encoding strategy, and need for pilot synchronization play a significant role on the effective user transmission data rate  \cite{meoni2020ccsds}. 

However, high-fidelity modelling of these factors can be tremendously complex requiring dedicated simulation software, such as \textit{ns-3} or \textit{OMNeT++} \cite{riley2010ns,varga2010overview}. 
In the long term, PASEOS may possibly provide interfaces for these packages. 
For the moment, however, users are required to manually specify an average data transmission rate to be considered for computing required transmission times, which is considered constant and, within PASEOS, independent of the distance between the actors and the communication link.

\subsubsection{Power Budgets}
Another constraint relevant for all kinds of space missions are the available power budget and power systems \cite{furano2020towards, FPGAOnboardICASSSP2019}.

Near the Sun and inner planets, spacecraft typically rely on solar power and batteries. Further out, spacecraft rely on radioisotope thermometric generators  \cite{patel2004spacecraft}. 
Land-based assets such as rovers typically also rely on either of these methods. 

We assume spacecraft to have a battery with a fixed capacity. For simplicity, we model solar power generation in PASEOS as an increase in the battery's state of charge (SoC) at a constant rate while a spacecraft is not in the eclipse of the central body relative to the Sun. 
For ground-based actors, power budget is currently not considered as we are focusing mostly on Earth-based ground segments. 
However, modelling of ground-based assets' power budgets can follow analogously to spacecraft if, e.g., rover operations are modelled. 

The SoC is reduced through the spacecraft activities, which---in the context of PASEOS---are user-defined operations that consume charge at a constant rate during operations.

\subsubsection{Thermal Management}
Another critical constraint on the activities a spacecraft can perform is posed by managing the spacecraft's temperature, both in terms of low and high temperatures. 
Spacecraft hardware will typically have a limited operational and survival range in terms of temperatures \cite{wertz1999space}. 
Hence, it is important that such constraints are respected during any operations the spacecraft performs. 
Due to the surrounding near-vacuum, heat dissipation is typically a major concern. 
However, low temperatures can also pose problems, especially when the spacecraft is in the central body's eclipse or deep space. 
Therefore, it is necessary to rely on a thermal model to anticipate and prevent violation of these constraints. 

In PASEOS, we rely on a formulation similar to the one described by Martínez \cite{martinez1995spacecraft}. We model the change in the spacecraft's temperature $T$ as
\begin{equation}
    mc \frac{\Delta T}{\Delta t} =  
    \delta_\mathrm{a} \dot{Q}_{\mathrm{solar}} +  \delta_\mathrm{a} \dot{Q}_{\mathrm{albedo}} +
    \dot{Q}_{\mathrm{IR}} + \dot{Q}_{\mathrm{activity}} - \dot{Q}_{\mathrm{diss}} 
    \label{eq:thermal}
\end{equation}
where $m$ is the spacecraft mass, $c$ its thermal capacity, $\delta_\mathrm{a} = 1$ if the spacecraft is in eclipse and 0 otherwise and the $\dot{Q}$ are heat fluxes. 
In particular, the individual heat fluxes are given as
\begin{equation}
    \dot{Q}_{\mathrm{solar}} = \alpha_\mathrm{a} \: E_\mathrm{s} \: A_{\mathrm{a, Sun}}
\end{equation}
where $\dot{Q}_{\mathrm{solar}}$ is the radiative flux produced by the sun, $\alpha_s\in[0,1]$ is the spacecraft solar absorptance, $A_{\mathrm{s, Sun}}$ is the area facing the Sun and $E_\mathrm{s}$ the solar irradiance.
\begin{equation}
    \dot{Q}_{\mathrm{albedo}} = 0.5 \: \alpha_\mathrm{a} \: \rho_\mathrm{b} \: E_\mathrm{s} \: A_{\mathrm{a,albedo}}
\end{equation}
where $\dot{Q}_{\mathrm{albedo}}$ is the sun heat flux generated back-scattered by Earth, $A_{\mathrm{a, albedo}}$ being the area facing the albedo and $\rho_\mathrm{b}$ being the central body's reflectance. 
The $0.5$ factor stems from a simplification as we currently do not compute angles between the Sun, the central body and spacecraft to reduce computational costs. 
\begin{equation}
    \dot{Q}_{\mathrm{IR}} = \frac{r^2_\mathrm{b} \: \epsilon_\mathrm{a} \: \epsilon_\mathrm{b} \: \sigma \: T_\mathrm{b}^4 \: A_{\mathrm{a,b}}}{\bar{r}_\mathrm{a}^2}
\end{equation}
where $\dot{Q}_{\mathrm{IR}}$ is the radiative flux due to the Earth black body emission, $A_{\mathrm{a,b}}$ is the spacecraft area facing the central body, $\bar{r}_\mathrm{a}$ is the spacecraft's distance to the central body center, $\epsilon_\mathrm{a}\in[0,1]$ the spacecraft's infrared emissivity (i.e. absorptance), $\sigma$ the Boltzmann constant and  $r_\mathrm{b}$, $\epsilon_\mathrm{b}$ and $T_\mathrm{b}$ are the central body's radius, infrared emissivity and temperature. Furthermore,
\begin{equation}
    \dot{Q}_{\mathrm{activity}} = \kappa P_\mathrm{A}
\end{equation}
where $\dot{Q}_{\mathrm{activity}}$ is the heat flux from the spacecraft hardware, $P_\mathrm{A}$ is the power consumption rate of an activity $A$ and $\kappa$ is a user-defined parameter describing conversion rate into heat. And finally, 
\begin{equation}
    \dot{Q}_{\mathrm{diss}} = \epsilon_\mathrm{b} \: A_\mathrm{b}\:\sigma T^4
\end{equation}
where $\dot{Q}_{\mathrm{diss}}$ is the heat emitted from the spacecraft.
While PASEOS does not enforce consideration of the spacecraft temperature (except disallowing temperatures below 0K) it enables the user to query the current temperature and formulate abort conditions based on it.

\subsubsection{Radiation Effects}
Another constraint and physical factor to consider are the effects of radiation, especially beyond LEO, where spacecraft are still fairly protected by the Earth's magnetic field \cite{srour1988radiation, lu2019review}. In practice, these events can lead to data corruption, software faults, or permanent hardware damage \cite{bruhn2020enabling}. 
In PASEOS, we model three different types of effects on operations due to single event effects (SEEs) \cite{furano2020towards}:
\begin{itemize}
    \item \textbf{Data corruption} with a certain probability due to single event upsets. For that, PASEOS models flipped bits that occurs according to a Poisson-distribution with a rate $r_\mathrm{d}$
    \item \textbf{Unexpected software faults} leading to a random interruption of activities following a Poisson distribution with rate $r_{\mathrm{i}}$
    \item \textbf{Device failures} following a Poisson-distribution with rate $r_\mathrm{f}$, which can be imputed mostly to single event latch-ups
\end{itemize}
Given the dependence on spacecraft-specific hardware and orbit, the definitions of these rates are left to the user.

\subsubsection{Operational Constraints}
Finally, in addition to the constraints imposed due to physics and the space environment, missions typically have many objectives and there are various stakeholders for any spacecraft \cite{grasset2013jupiter}. Therefore, there are often operational constraints to be considered that are imposed through the mission profile. PASEOS enables these by allowing user-defined constraints based on arbitrary parameters that are evaluated during activities and lead to interruption of the activity if the constraints are not met. This gives users a broad range of possibilities ranging from imposing strict hardware limits, such as having to respect a minimum SoC or certain temperature range, to requiring time limits or factors outside the PASEOS simulation.

\subsection{Software Design}
In the following section, we briefly elaborate on the design philosophy of PASEOS with regard to user interaction and validation of the models. 
Even though we already provide concrete application results in this work, the aim of PASEOS is to enable a variety of future applications and consequently a flexible and generic design is paramount.

\subsubsection{Actors}
As an abstraction of the variety of different assets available on ground and in space we simulate them in PASEOS as so-called \textit{actors}. 
Fundamentally, we distinguish between ground-based actors and spacecraft actors. 
In the current version, the main feature associated with the former is modelling the change in position due to the rotation of the central body. 
For spacecraft actors, a variety of models can be enabled describing all the physical and operational aspects described in Sec.~\ref{sec:constraints}. 
Initially, one only needs to define the type of actor, a name for it and the current local time of the actor. Depending on the desired models, additional parameters, e.g. position and velocity for orbits, need to be specified.

\subsubsection{Activities}\label{sec:activities}
Activities are the second central abstraction in PASEOS. They serve to describe any kind of operation the user wants to model. For example, we may want our spacecraft actor to capture data with one of its sensors and process that data. PASEOS allows the specification of an (asynchronous) function that is executed in the background while PASEOS models the physical constraints. Further, a constraint function, which is then evaluated repeatedly at a fixed timestep, can be used to interrupt an activity. 
For each activity, one has to specify the power consumption to allow PASEOS to model excess heat and the change in the battery's SoC. 

Users can either let PASEOS run operations asynchronously to model updates or run operations and then advance PASEOS' simulation time.

\subsubsection{Discrete-Event vs. Time-based}
One additional challenge for PASEOS is that network-oriented simulations, such as \textit{ns-3} \cite{riley2010ns} typically focus on providing a discrete-events simulation allowing users to skip to the next event. Physical simulation modelling ordinary differential equations, such as the one in Equation \ref{eq:thermal}, however, are usually solved in a time-based fashion with discrete time-stepping schemes \cite{ozgun2009discrete}. In PASEOS, these two simulation paradigms meet as both kinds of simulations are addressed and part of PASEOS. 
PASEOS operates at the intersection of both as it performs numerical simulations of physical processes while modelling discrete events such as communication windows becoming available. 
One can choose a fully real-time operation of PASEOS, where it will run the physical models asynchronously while performing a user-defined activity. 
Alternatively, one can manually ask PASEOS to advance its simulation state to a specific time interrupted by potential events of interest to the user occurs. 

\subsubsection{Using PASEOS}
\begin{figure}[!htbp]
\centering
\includegraphics[width=\linewidth]{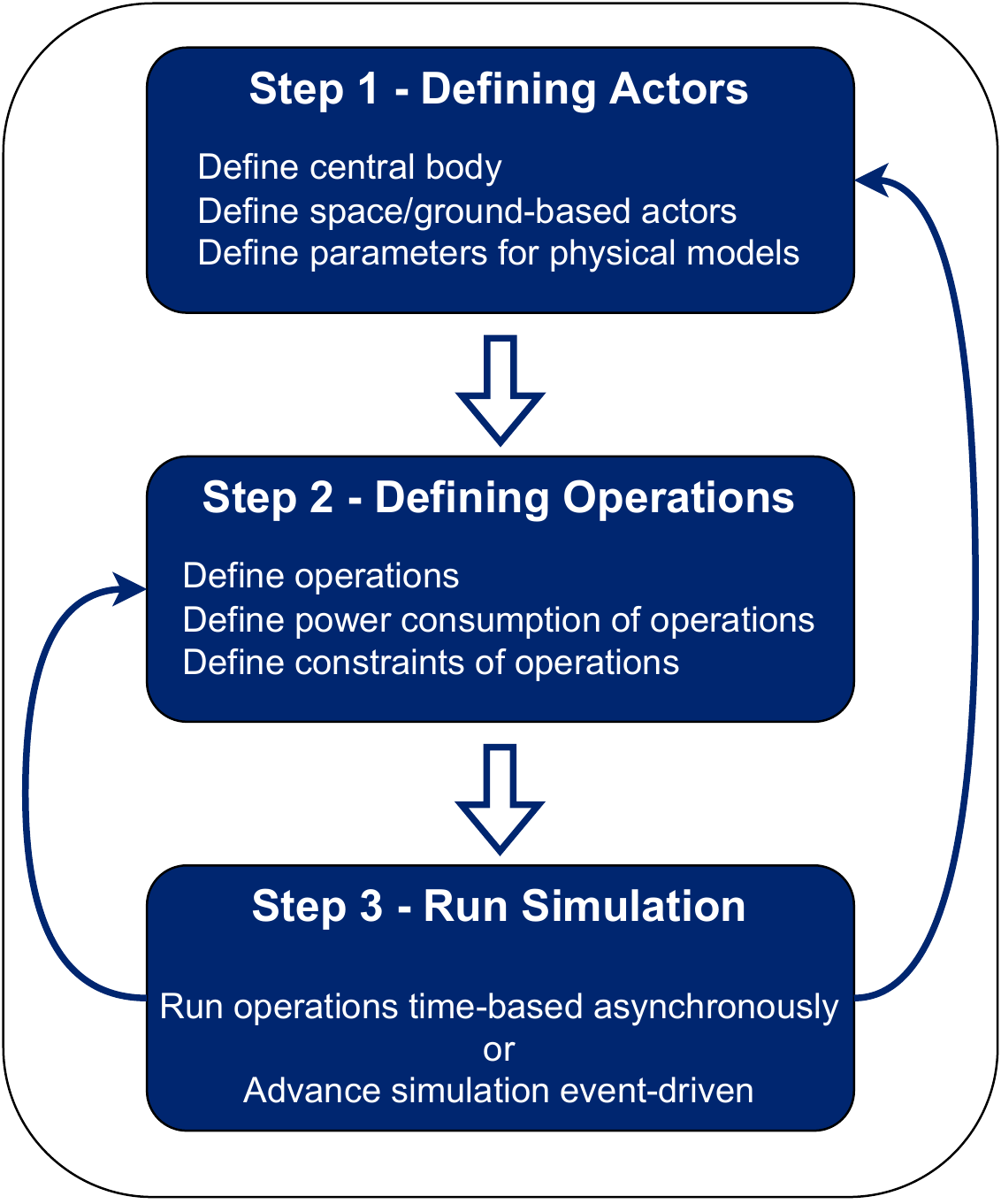}
\caption{Overview of the workflow using PASEOS \label{fig:workflow}}
\end{figure}

With the relevant terminology introduced, we can describe the main workflow of PASEOS. Figure \ref{fig:workflow} gives an overview of the high-level use of PASEOS. It fundamentally requires two definition steps where, first, the actors are modelled and, second, the modelled operations are defined. 
Operations can then be performed repeatedly in either the discrete-event or time-based fashion. 
At any point of the simulation, the user may benefit from detailed output logs on actor status and activities (written to a *.csv file) and/or from visualizations.

\subsubsection{Validation}
Given the multitude of physical models in PASEOS as well as the complex interaction due to asynchronously running activities, a thorough validation of the software is both critical and challenging. 
We rely on a comprehensive suite of automated tests to validate the continued correctness of the implemented models. 
In many cases, the tests are based on realistic scenarios, such as Sentinel-2 communication windows. 
Test-driven development has been employed for the design of many of the individual components and no contributions are merged without code review, appropriate tests, and passed automated tests. 
A modular design approach enables testing of individual components, such as the thermal model or communication links. 
Further, PASEOS is available online as an open-source software to enable anyone to provide feedback and report bugs.

\subsection{Modelling Asynchronous Operations between Multiple Spacecraft}

The PASEOS framework may be readily used to model operations involving multiple actors where each physically modelled actor contains a PASEOS instance.
Note that this setting requires a change of perspective from Fig.~\ref{fig:workflow}, where there is a single actor of interest, to a system-level view where each actor is to be modelled. 
PASEOS does not aim to facilitate coordination, scheduling of activities, or other tasks specific to the modelled application. 
Instead, it aims to provide the user with the capability to model different scenarios, such as intermittent knowledge of other actors and both centralized (e.g. federated learning) and decentralized (e.g. decentralized learning) operations.
Each actor is assumed to operate independently and any information about other actors should be acquired during the operation. 
Operations can either be performed independently, where the user takes the responsibility to advance the time of the PASEOS simulation manually, or they can operate in an asynchronous fashion with event-based activities. 

Flexibility is an important design goal of PASEOS, and the user is able to equip a PASEOS instance with arbitrary capabilities by registering activities, see Sec.~\ref{sec:activities}, consisting of the actions to be performed, the power consumption, and constraints that must be satisfied for the activity to be performed.
The action and constraint of an activity are coroutines, i.e., asynchronous functions, and rely on the \textit{asyncio} Python module \cite{asyncio}. 
When an activity is performed, its action and constraint function are submitted to an event loop for execution.
Meanwhile, PASEOS will monitor and update the internals of the actor, e.g., temperature and SoC, and advance the time.
Note that PASEOS allows only one activity to be performed simultaneously. 
Given the arbitrary code execution inside the activity, a user can, however, perform different tasks within an activity, e.g. based on the actors state.

To allow for simulation of multiple spacecraft, the ability for the actors to interact is imperative.
In PASEOS, communications are achieved by encapsulating transmission and reception as activities.
Such activities should be designed to comply with the communication device for the given actor, see Sec.~\ref{sec:tx_bw}, in order for the communication windows to be properly calculated.
For simulation on a single machine, communications may be emulated by simply imposing a delay (and power consumption) in the corresponding activity.

To use PASEOS directly on edge devices, e.g., spacecraft hardware, communication activities may be based on packages such as \textit{gRPC} and \textit{zeroMQ} to allow interaction over a network.
When working with edge devices, time synchronization between the actors is paramount for correct operation.
As the actors in PASEOS operate independently, they must make themselves known to others by, e.g., sending out heartbeats at given intervals containing the id of the actor and, optionally, metadata, e.g., position and velocity. 
An actor may also be unavailable due to low SoC, no line-of-sight connection, or a device failure.
The absence of heartbeats entails the unavailability and other actors may respond accordingly. PASEOS does not automatically send heartbeats or check these factors but provides an API to serialize actors for network transmission and the user to tell each actor about its known and available peers. Thus, arbitrarily simple or complex connectivity constraints can be imposed.

If one wants to perform specific operations in a synchronized manner it is necessary to synchronize actors in time by, e.g., providing a start time for an activity.
Alternatively, time synchronization can be achieved by means of communication~\cite{arenas2008sync}.
For example, one may utilize the Network Time Protocol~\cite{mills1991ntp} that relies on a master clock or decentralized approaches that rely on, e.g., the heartbeats~\cite{babaoglu2007firefly}.

\section{Results}\label{sec:results}

In this section, the usage of PASEOS is demonstrated for three distinct modelling scenarios: Earth observation, constellation design, and decentralized machine learning.
The purpose is to illustrate the broad range of applications that are readily modelled with the aid of PASEOS.

\subsection{Single Actor: Onboard Volcano Detection Onboard Sentinel-2}\label{sec:profiling}
This experiment showcases how one can use PASEOS in real-time simulations to emulate satellite onboard-processing scenarios on actual space hardware. 
\subsubsection{Setup}

We design an experiment in which we register an activity to process satellite payload data and utilize a constraint function to check the availability of a link to transmit data to the ground. 
During the simulation, we profiled our code on a Unibap iX-10 100 satellite processor \cite{bruhn2020enabling} to evaluate the overhead of PASEOS, i.e., the time spent to update the physical models compared to the time required for checking constraints and running user activities.  
In particular, we consider onboard detection of volcanic eruptions applied to Sentinel-2 L1-C where the acquired data are processed directly onboard a satellite to spot possible volcanic eruptions and deliver early-alerts to ground \cite{del2021board, di2022early}.

 To set up the orbit of the spacecraft actor we used the ephemerides of the Sentinel-2B satellite at 2022-10-27T12:30:00Z, which were calculated by using two-line elements. 
 Because of that, the orbit of our actor is sun-synchronous. 
 To set up the ground station actor position, we used the European Space Agency ground station, located at Maspalomas, Gran Canaria (27.7629\degree{} latitude, -15.6338\degree{} longitude at an elevation of 205.1m). The ground station link is available if the satellite is 5\degree{} above the horizon. 
 To check for the possibility to communicate, we used a constraint function that interrupts the user activity (i.e. the volcanic eruption detection) when the satellite is in line of sight (LOS) with the ground station.
 
To have sufficient energy to process data onboard, the actor is equipped with a 0.162 MJ battery, with a SoC of 1.0 at the beginning of the simulation. 
We assume a charging rate of 10 W.
Furthermore, to increase the computational cost due to the update of the physical models in PASEOS and test the system in the worst case, we also equipped the space actor with a thermal and radiation model to measure their run time impact. 
In particular, to avoid effects of radiation but track computational cost, we set up the rate of data corruption, restart, and device failure to be 0---this does not affect the computational cost. 

The simulation data consist of three Sentinel-2 L1C post-processed products showing volcanic eruptions of Etna (2021-08-30), La Palma (2021-09-30), and Mayon (2018-02-09), provided by Meoni et al. \cite{meoniLPS} \todo{@Gabi update}.
Each image is produced by cutting and mosaicing the 20m bands B8A (Near-Infrared) and the  B11 and B12 (Short-Wave Infrared) of multiple tiles over the area of the band B8A of the correspondent Sentinel-2 L0 granule \cite{S2PSD}.
We artificially extended the simulation time by repeatedly evaluating the images until we get 100 images in total for the simulation. In addition, we assumed all data to be already acquired and available for processing to disregard hardware-related delays that would occur in reality as we focus here on profiling PASEOS.

Code profiling was performed by using the Python module \textit{yappi} \cite{yappi}. In particular, we measured the Central Processing Unit (CPU) time on the iX-10 100 device for three different values (0.25 s, 0.5 s, 1 s) of the \textit{PASEOS timestep}, i.e. the interval time for updating PASEOS, its physical models, and checking user constraints.
We performed three runs for each choice of PASEOS timestep. 
During each run, we measured the CPU time for the user activity, to check user constraints (i.e., check of LOS with the ground station), time to update PASEOS overall, and the individual times spent to update the radiation, the thermal, and the battery charge models. 
Results were averaged over the three runs for each test case.
All the tests were performed sequentially with two warm-up runs with PASEOS timestep of 1 s. 

\subsubsection{Onboard volcanic eruption detection}
\begin{figure}[!bp]
\centering
\includegraphics[width=\linewidth]{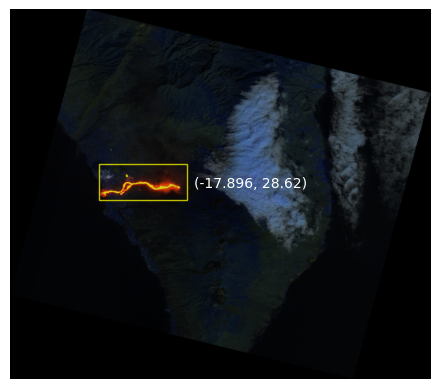}
\caption{Detected volcanic eruption on the La Palma (Spain) on 2021-09-30. The displayed coordinates (longitude, latitude) correspond to the center of the detected bounding box.}

\label{fig: EruptionDetection}
\end{figure}

The detection is based on a simplified implementation of the algorithm~\cite{massimetti2020volcanic}, presented in~\cite{meoniLPS}, \todo{@Gabi update}which produces a bounding box surrounding the detected volcanic eruption and the associated geographical coordinates. Since the aim of the activity is to provide an early alert to promptly notify and locate a detected volcanic eruption, a possible alert message will provide the coordinates of the bounding box center or top-left/bottom-right points.
One example of a detected volcanic eruption on the island La Palma is shown in Fig. \ref{fig: EruptionDetection}.

\subsubsection{Results of profiling}
\label{subsubsec: ProfilingResults}

Tab. \ref{tab: IX10Results} shows the simulation results for the runs on the iX-10 100 device. 
In each test, the activity was interrupted after roughly 29 s of CPU time because the satellite was found in LOS with the ground station. Indeed, the time to perform the user activity is similar for each value of the PASEOS timestep, whilst the time to update PASEOS state and to check the user constraints grow almost linearly. However, even for the smallest PASEOS timestep, the latter are respectively 0.43 s and 0.29 s compared to 28.64 s for the user activity. Overall, PASEOS requires around the 1.45\% of the total run time.  This demonstrates that PASEOS is a lightweight solution that can be suitable for onboard processing use cases on embedded hardware. 

As can be seen in Tab. \ref{tab: IX10Results} the time for modelling battery charge is equal to the one for updating the thermal model. 
This is because the update of both the power and the thermal models require checking whether the spacecraft actor is in eclipse. 
This operation is carried out only once for both the models and run time of it is attributed equally to the thermal and power models. All the other operations to update the battery SoC and thermal models are negligible.

\begin{table}[tb]\scriptsize
\caption{Results of profiling on a Unibap iX-10 100 device for different PASEOS timesteps. Each column reports the CPU time in seconds spent for the user-defined activity, to update the physical models of PASEOS and their individual times. Percentages with respect to the total time are reported in brackets.
The last column provides the time spent for the check of the LOS.}
\label{tab: IX10Results}
\begin{center}
 {\tabulinesep=1.4mm
 \setlength\tabcolsep{2pt}
\begin{tabu} {|  X[0.625cm]|  X[0.75cm]||  X[1.0cm]| X[1.0cm]| X[1.0cm]|}
\cline{1-5}
\multicolumn{2}{|c||}{All in [s](\%)}&  \multicolumn{3}{c|}{PASEOS timestep} \\ \cline{1-5}
 \multicolumn{2}{|c||}{Type}                                & 0.25              & 0.5           & 1.0 \\ \hline \hline
 \multirow{2}{*}{\rotatebox[origin=c]{90}{\parbox{1cm}{User-\\defined}}} & User Activity & 28.64 (97.47)     & 28.59 (98.66) & 28.52 (99.27) \\ \cline{2-5}
 & LOS Constraint                                           & 0.29 (0.98)       & 0.14 (0.50)   & 0.072 (0.25)  \\ \cline{1-5}
 \multirow{5}{*}{\rotatebox[origin=c]{90}{\parbox{2cm}{PASEOS \\ State Update}}} & SoC  & 0.11 (0.39)       & 0.062 (0.21)  & 0.034 (0.12) \\ \cline{2-5}
 & Radiation                                                & 0.0086  (0.029)   & 0.0044 (0.015)& 0.0024 (0.008) \\ \cline{2-5}
 & Thermal                                                  & 0.11 (0.39        & 0.062 (0.21)  & 0.034 (0.12) \\ \cline{2-5}
 & Other                                                    & 0.19    (0.65)    & 0.10 (0.34)   & 0.056 (0.19) \\ \cline{2-5}
 & Subtotal                                                 & 0.43 (1.45)       & 0.23 (0.79)   & 0.13 (0.44) \\ \cline{1-5}
 \multicolumn{2}{|c||}{Total}                               & 29.39             & 28.98         & 28.74 
 \\ \cline{1-5}
\end{tabu}}
\end{center}
\end{table}

\subsection{Multiple Actors: Communications Modeling of a Constellation} \label{sec:constellation}
The capabilities of PASEOS to model the operational constraints of managing a constellation are demonstrated in this test case. Detailed results on the constellation's status over time are given and a simple scaling study is performed to investigate PASEOS scaling abilities to showcase the potential of modeling large constellations.
\subsubsection{Setup}
The scenario investigated here is a LEO constellation consisting of sixteen spacecraft in a Walker pattern \cite{walker1984satellite} in four planes at 550 km altitude with an inclination of 10\degree. Operations of the constellation are modelled for eight hours---that equals slightly more than five revolutions. The satellites are presumed to be equally equipped with a 1 MJ battery initially at randomly uniform SoC between 0.1 and 1.0. Each satellite is equipped with solar panels charging at 50 W. Satellites are assumed to have a mass of 50 kg, to be at 273.15 K initially and have absorptance of solar and infrared light of 1.0. The areas facing the Sun and Earth are assumed to each be 2 m². The emissive (radiating) area is presumed to be 4 m². The thermal capacity is assumed to be 1000 Jkg$^{-1}$K$^{-1}$. We assume half of the used wattage for satellite operations to be converted to heat. The Earth's temperature is assumed to be 288 K, its infrared emissivity at 0.6 and its solar reflectance to be 0.3. Solar irradiance is estimated as 1360 W. 

Satellites have two operational modes, one is a standby mode called \textit{Standby} where they consume only 2 W and the other one is called \textit{Processing} where they consume 100 W. 
The satellites automatically switch to \textit{Standby} if their battery's SoC falls below 0.2 or their temperature above 330 K. 

In addition to the constellation, we monitor availability of communication links to a satellite in geosynchronous orbit called \textit{GeoSat} and the European Space Agency ground station at Maspalomas, Gran Canaria. As for the single satellite scenario, the ground station link is available if satellites are 5\degree{} above the horizon. The geosynchronous satellite is reachable from the Maspalomas station. 

In PASEOS, we use a time step of 1 s for the thermal and power models. Satellites decide every 600 s whether they are ready for \textit{Processing}. If the constraints for \textit{Processing} are violated during the 600 s interval, they switch to \textit{Standby} for the remainder of the interval. 

\subsubsection{Constellation Analysis}
\begin{figure}[!htbp]
\centering
\includegraphics[width=\linewidth]{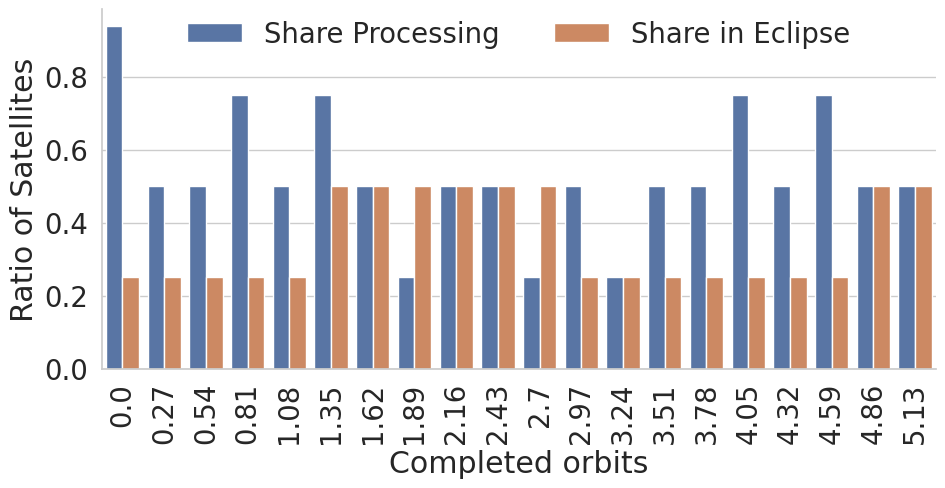}
\caption{Overview of the satellites able to process and in eclipse over time. \label{fig:constellation_active}}
\end{figure}

\begin{figure}[!htbp]
\centering
\includegraphics[width=\linewidth]{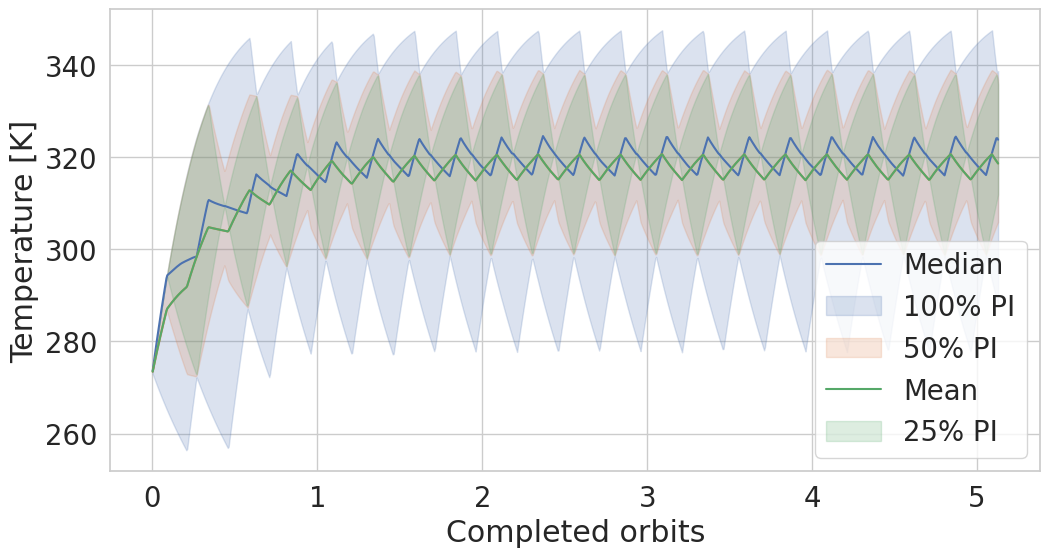}
\caption{Temperature of the constellation. Different colors indicate percentage intervals around the median. \label{fig:constellation_temp}}
\end{figure}

We analyze the constellation regarding several operational factors. First off, in terms of time spent processing. As can be seen in Fig.~\ref{fig:constellation_active}, roughly half of the satellites are processing at any time, and both battery SoC and temperature limit the periods of operation. Given the circular LEO orbits of the constellation, the satellites spend a large time in eclipse with 25 to 50\% of the constellation being in eclipse at any moment. This also influences the operational temperature---as seen in Fig.~\ref{fig:constellation_temp}---of the satellites which rises quickly at the beginning when a large share begins processing and the temperature falls especially during eclipse. 

\begin{figure}[!htbp]
\centering
\includegraphics[width=\linewidth]{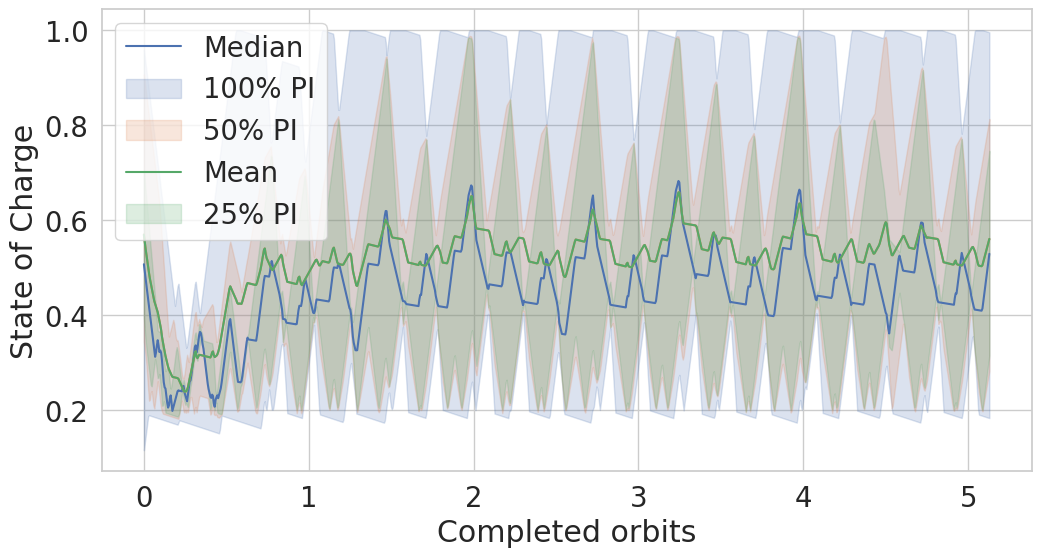}
\caption{Battery SoC of the constellation. Different colors indicate percentage intervals around the median.\label{fig:constellation_power}}
\end{figure}

Even though PASEOS' SoC model, for now, is simplistic, complex dynamics for the power budgets can be observed in Figure \ref{fig:constellation_power}. Satellites in the constellation fluctuate between a SoC of 0.2 and 1.0. The low standby consumption of the \textit{Standby} activity means they never run the risk of reaching critical SoCs. Overall, the constellation's behavior in terms of satellites performing \textit{Processing} and the constellation's temperature and SoC become stable and cyclic after roughly one orbital revolution. 

\begin{figure}[!htbp]
\centering
\includegraphics[width=\linewidth]{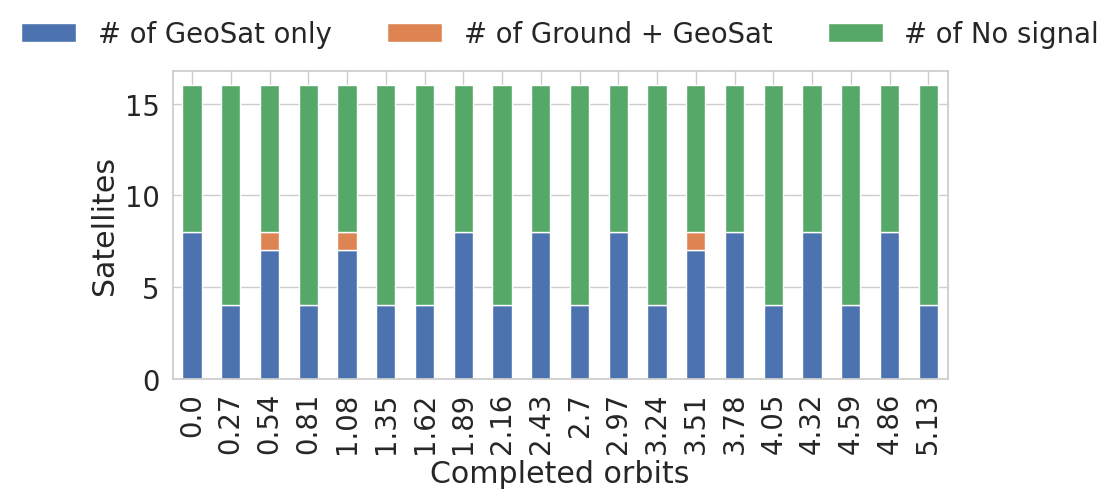}
\caption{Overview of the satellites' communication status over time. \label{fig:constellation_comms}}
\end{figure}

In terms of communication status, 50 to 75\% of the constellation are not within LOS of either the ground station or satellite in geosynchronous orbit as can be seen in Figure \ref{fig:constellation_comms}. As the geosynchronous satellite is reachable from Maspalomas, it is also in LOS of a constellation satellite when the ground station is.

\subsubsection{Performance and Scaling}
Running this scenario on a AMD Ryzen-5 consumer-grade CPU on a single thread requires about 200 s. 
Time is spent almost exclusively on the physical models and LOS checks. 
With 16 satellites, running the simulation for a simulation time of 600 s requires 8.3 s, 0.52 s per satellite. 
We investigated the scaling by increasing the number of satellite per plane. 
With a constellation of size 32, 16.4 s were necessary for 600 s simulation time, 0.51 s per satellite. 
The better-than-linear scaling is likely due to a constant initialization overhead in starting the simulation. 
With 512 satellites, the 600 s required 255.9 s computation time or 0.50 s per satellite. 
Thus, it is clear that PASEOS scales well even to a large number of satellites. 
Parallelization of these simulations is also trivial given the fully asynchronous nature of PASEOS that permits parallelization without any concern for shared memory of similar.

\subsection{Onboard Machine Learning in Orbit} \label{sec:onboard_learning}
This test demonstrates how PASEOS can be employed to model and monitor constraints when solving a machine learning task in a decentralized setting. Limited communication windows, heterogeneous data, and power constraints are imposed while successfully solving a classification task.

\subsubsection{Setup}
The operational scenario in this example consists of two satellites in circular LEO at an altitude of 550 km with an inclination of 98.62\degree. They move in opposing directions and thus have only brief communication windows amongst them twice per orbital revolution. Both are equipped with a 0.1 MJ battery with an initial SoC randomly chosen between 0.6 and 0.8. They have solar panels which charge the battery at 50 W when not in eclipse. They are also equipped with inter-satellite links enabling them to transmit 1 Mbit per second amongst themselves when in LOS. 
The satellites are assumed to not communicate during the first ten orbital revolutions.

\begin{figure}[!htbp]
\centering
\includegraphics[width=\linewidth]{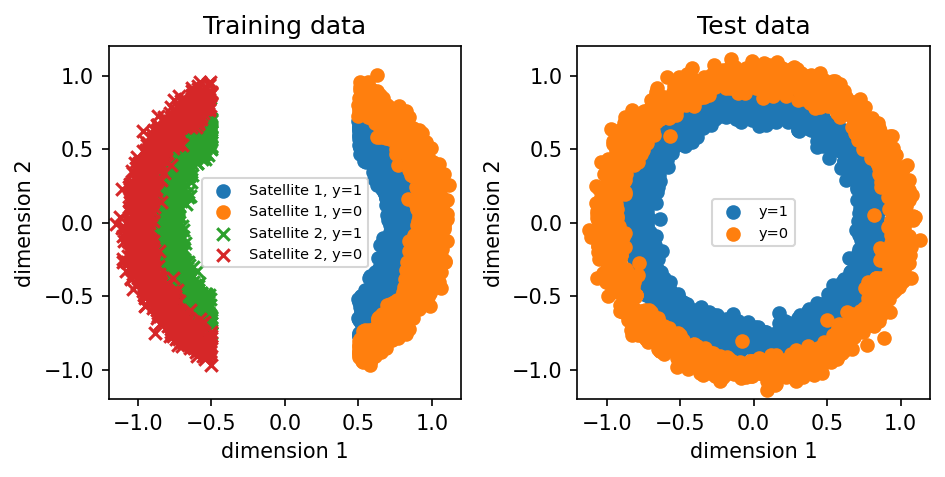}
\caption{Training and test data for the binary classification task as well as their distribution among satellites. Note the heterogeneity of the data distribution in the training set. $y = 0$ and $y = 1$ are the classes of the problem. \label{fig:learning_data}}
\end{figure}

The satellites are tasked with jointly learning a binary classification task identifying an inner and outer circle by leveraging data uniquely available to each satellite. 
The used dataset and its distribution among the satellites is depicted in Fig. \ref{fig:learning_data}. 
Notably the distributions are heterogeneous with the satellites only having access to data points with values in the first dimension above 0.5 and below -0.5, respectively.
The test dataset is identical on both satellites and covers the complete feature space. A total of 4166 training samples are used and the test dataset consists of 3300 samples. A two-layer dense neural network with ten neurons is trained using stochastic gradient descent with a learning rate of 0.1 and a binary cross-entropy loss function. Training is modelled for a total of 30 orbital revolutions. 
In each revolution, if allowed, the satellites communicate twice and will train an aggregated model in the window between communications. 
On average, Satellite 1 is able to train 42.3 epochs and 11.6 epochs in the two different windows whereas Satellite 2 is able to train 4.1 epochs and 50.5 epochs.
During operations, the satellites have three distinct operational modes they perform in descending priority:
\begin{itemize}
    \item \textbf{Standby} -- When the SoC is below 0.5 the satellites stand by to conserve and charge their batteries. This mode drains 2 W.
    \item \textbf{Model Sharing} -- When the satellites are in LOS of each other they exchange their models and aggregates via a federated averaging at a power consumption of 100 W.
    \item \textbf{Model Training} -- If neither of the above takes place the satellites perform a training epoch at a power consumption of 100 W.
\end{itemize}

\subsubsection{Learned Model and Communications}

\begin{figure}[!htbp]
\centering
\includegraphics[width=\linewidth]{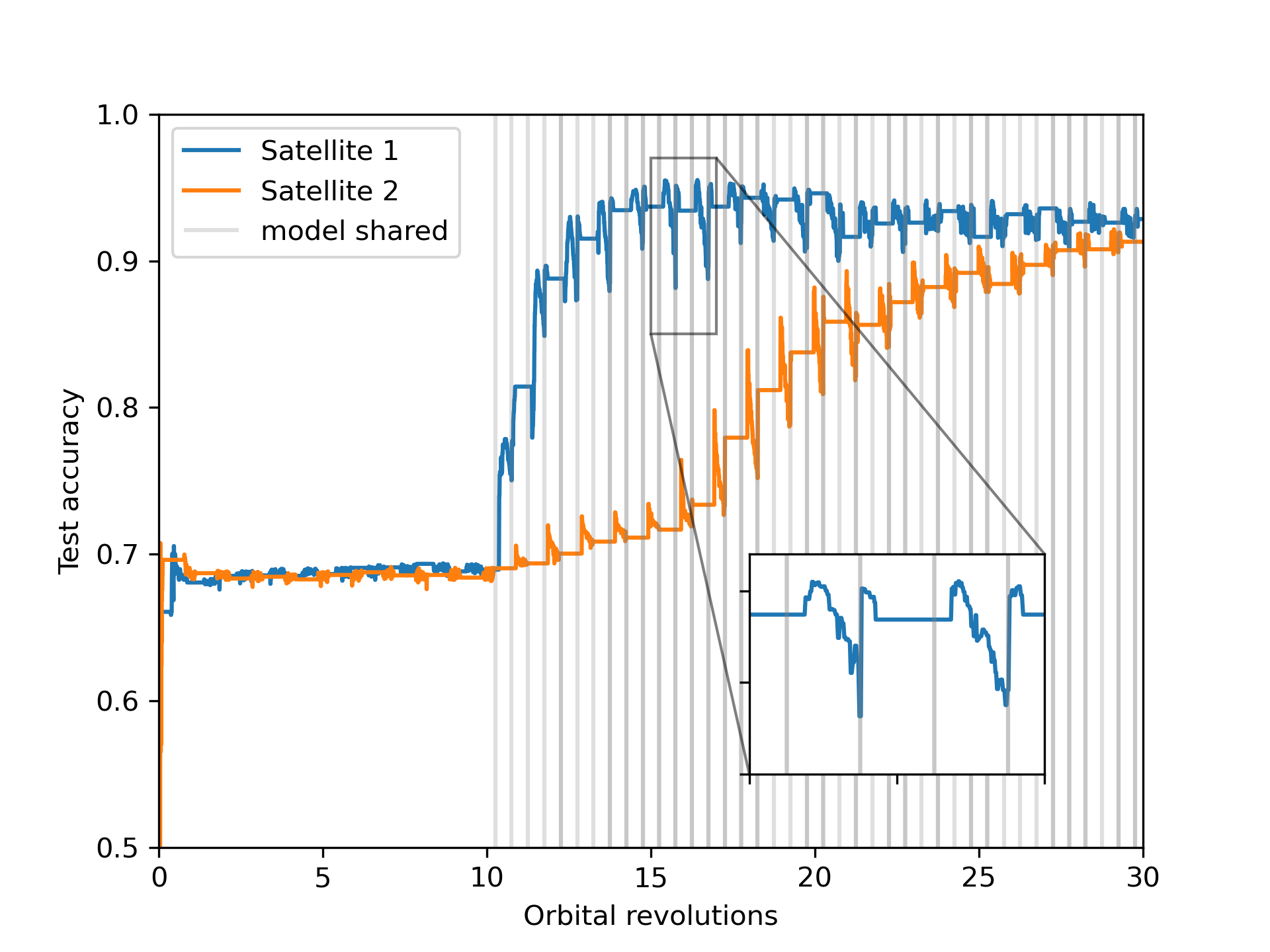}
\caption{Test accuracy over 30 orbital revolutions. Vertical gray lines indicate communication between the satellites. Constant accuracy stems from the battery SoC being below 0.5. \label{fig:learning_acc}}
\end{figure}

Overall, training the models on the two satellites is successful, reaching an accuracy of over 91\% on both satellites compared to a random guess accuracy of 50\% and independent training that results in an accuracy below 70\%. Figure \ref{fig:learning_acc} displays the test set accuracy over time and shows how both satellites struggle to obtain good results individually. However, when they exchange models (marked with gray vertical lines) they improve their accuracy noticeably. 
Notice that model transmission is virtually instant given the small models (only 1312~bits).

It can also be seen that each satellite benefits after each communication round as the test accuracy rapidly increases. 
However, once the satellite starts training on local data again, the performance deteriorates due to catastrophic forgetting~\cite{shoham2019forgetting}.
Furthermore, there is a large discrepancy between the test accuracy for the two satellites. 
This happens as Satellite 1 is able to do training after each model exchange because charging is initiated after one exchange and the satellite remains in LOS with the sun after the next exchange.
Satellite 2, on the other hand, is charging before one exchange and has stopped charging during the next.
Hence, Satellite 2 may not have battery to do training after one of the model exchanges and, therefore, its convergence is slower, as seen in Fig.~\ref{fig:learning_acc}.

\begin{figure}[!htbp]
\centering
\includegraphics[width=\linewidth]{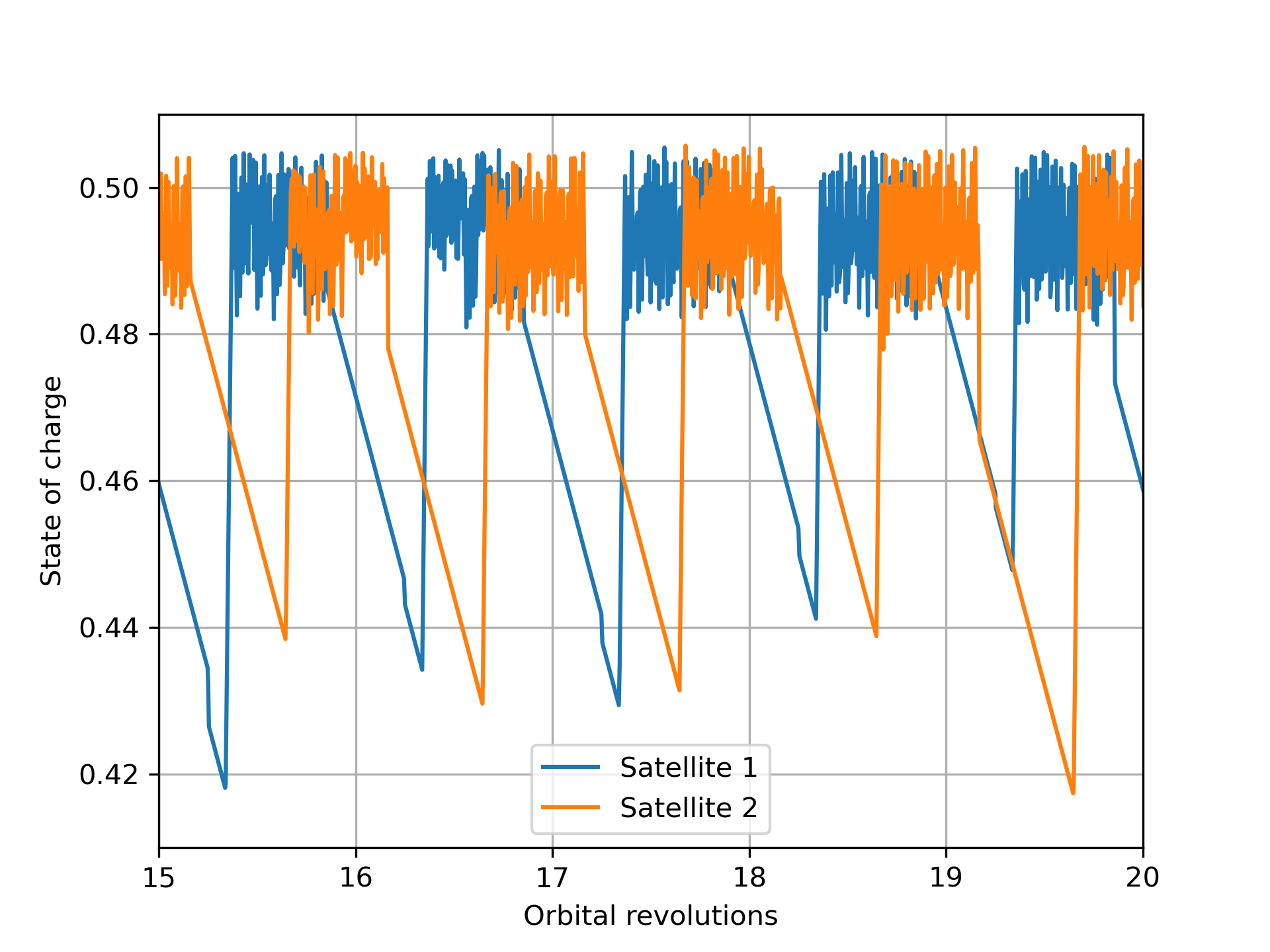}
\caption{SoC between orbital revolution 15 and 20.} \label{fig:learning_soc}
\end{figure}

In terms of power consumption, which is displayed in Fig.~\ref{fig:learning_soc}, we can observe training in the intervals of rapid oscillation of the SoC when the battery is charged and consequently the training run. During eclipse the satellites go into standby to conserve SoC and drains only 2 W. Finally, the rapid increase in SoC occurs when a satellite is facing the sun and training is not allowed because SoC is below 0.5. 

\section{Discussion}

In general, the examples in Sec.~\ref{sec:results} clearly demonstrate the viability of PASEOS for the three considered test cases. 
By running on an actual satellite processor in real time, we have also shown the feasibility of using PASEOS to model scenarios while utilizing real, prospective mission hardware. 
In the case of the LEO constellation, we can see that PASEOS is also capable of modelling large constellations on a longer time frame. 
Finally, we also showcased how PASEOS can be used to model constraints that impact the training of machine learning models directly in space. 
There are, however, some aspects that mandate further discussion and consideration.

\subsection{Impact of Onboard Constraints on Machine Learning}
One of the objectives of PASEOS is to study the impact of operational constraints when utilizing machine learning methods in orbit. 
Especially in the context of distributed and onboard learning paradigms it is an essential question what this impact will be \cite{matthiesen2022federated,razmi2022board}. 
As demonstrated in our onboard learning results in Sec.~\ref{sec:onboard_learning}, PASEOS can provide concrete insights into the operational impact of factors such as orbital dynamics and power consumption. 
As can be seen in Fig. \ref{fig:learning_acc}, the timing of communication windows and eclipse directly influenced the accuracy obtained during the distributed training. 
While the importance of factors such as temperature, battery SoC, or the timing of communication windows in relation to eclipse are well established parameters for spacecraft operators they have not been studied in the context of the application of machine learning methods in orbit. 
Even in-orbit demonstrations, such as $\Phi$-Sat-1 \cite{giuffrida2020cloudscout}, did not explore the topic of continuous operations of these systems over a longer operational time frame where many of the constraints captured by PASEOS play an important role. 
For both inference and training in constellations, to the authors' knowledge, there is currently a lack of holistic modelling of these factors as offered by PASEOS. 

\subsection{Scalability \& Performance}
As shown on the single device example using real space-rated embedded hardware in Sec.~\ref{sec:profiling}, PASEOS is a lightweight tool that allows modelling the space environment and operational constraints with minimal overhead for the user activity. 
Results shown in Sec. \ref{subsubsec: ProfilingResults} showcase that the computational complexity inside PASEOS models scales linearly with the PASEOS timestep while requiring less than 1.5\% of the total CPU time compared to the user activity with the highest investigated update rate of PASEOS. 
This is due to the particularly efficient physical models that offer suitable trade-offs to be used on edge devices. This partially comes at the cost of model fidelity, for which the main limitations and possible future improvements are discussed in Sec. \ref{subsec: fidelity} and \ref{subsec: limitations}. 

On the other end of the simulation spectrum, the LEO constellation example in Sec.~\ref{sec:constellation} shows that even as PASEOS is able to handle large constellations of up to hundreds of satellites, performance does become a concern. If one wants to model the long-term viability of a constellation including aspects like station keeping, degradation of photovoltaic cells and batteries, and similar effects (which are yet to be implemented in PASEOS), computational time may become a concern. In the single-threaded, single-core run used for the example, the runtime would, at the moment, already be prohibitive for a constellation with hundreds of satellites on a timescale of months or years. Thus, a careful study of the parallelization possibilities of PASEOS will be required in the future. As PASEOS is implemented in a fully asynchronous manner, it is straightforward to employ tools such as the popular message passing interface (MPI) \cite{walker1996mpi} to use compute clusters for PASEOS---this is shown in one of the online examples. A more detailed investigation of this is warranted.

\subsection{Fidelity Considerations}
\label{subsec: fidelity}
One key point in terms of fidelity that has to be taken into account are the user-provided specifications. They are not studied here in detail as they are beyond the scope of this initial release. However, given the holistic nature and the complex emergent behavior---see, e.g., the complex SoC curve stemming from just a simple, linear charge model in Figure \ref{fig:constellation_power}---of PASEOS, the accuracy and quality of the input parameters is likely to become a critical factor in the fidelity of results produced by PASEOS. 
Notably, if all physical models are activated, the number of input parameters becomes fairly large. 
Indeed, just the thermal model requires eight parameters, the power model currently requires three, the radiation model another three, the orbital model seven and these parameters need to be defined for every single actor in PASEOS. 
Thus, even with the simplified models present in PASEOS, it is evident that constellation scenarios modelled with PASEOS are already highly complex systems. 
In the future, this may require more detailed analysis of the system sensitivity to specific parameters to guide users as to which parameters are particularly critical. 
It is also conceivable to add noise terms in a variety of the PASEOS's models to account for this sensitivity. Even higher robustness code could be achieved by converting some of the parameters to be described by distributions instead of scalar parameters and thus include a probabilistic, Bayesian modeling component that allows consideration of prior assumptions about the accuracy of passed parameters.

\subsection{Limitations \& Prospective Additions}
\label{subsec: limitations}
At the moment, there are some natural limitations to the fidelity of the models in PASEOS and supported scenarios. The two objectives of holistically modelling the operational environment in space and being able to execute PASEOS in the background in real-time on edge devices requires a careful balancing of computational cost and physical fidelity. In the future, this may be remedied by adding optional components to PASEOS that model the various physical aspects at a higher fidelity when enough computational resources are available. Initially, however, we have focused on breadth of considered physical aspects instead of high fidelity in individual aspects. 

In terms of astrodynamics modelling, Keplerian dynamics are a sufficient start for low-precision Earth orbits but insufficient to, e.g., model orbits around irregular bodies such as asteroids or comets. Similarly, phenomena such as station keeping and occasional losses of tracking cannot be model with them. Natural additions in the future would be a polyhedral gravity model \cite{tsoulis2012analytical} or the wrapping of a software like \textit{orekit} \cite{maisonobe2010orekit}. 

The availability of communication windows in PASEOS is currently determined solely by whether the LOS is being blocked by an assumed sphere with a specific radius. More complex geometry could be integrated via meshes. Other factors such as success of tracking, distance of the actors and atmospheric conditions for optical communications and similar are also currently not modeled but sensible additions. 
Similarly, the communication bandwidth that is currently available is in reality a more complex and variable quantity dependent on a lot of the just mentioned factors such as distance, link conditions, and others parameters linked to the transmission chain (i.e., channel-encoding, modulation, additional use of synchronization pilots, etc.). A more thorough channel modelling is required to account for this. A potential way forward is to wrap  \textit{ns-3} or \textit{OMNeT++} into PASEOS~\cite{riley2010ns,varga2010overview}. 

In regards to power budgets there also several conceivable improvements, such as more rigorous modeling of the state-of-charge of the battery \cite{lai2020parameter}, a more thorough model for the charging via solar panels \cite{porras2021simple} and consideration of factors such as the age and temperature of the battery, devices, and solar panels. 
The thermal model in PASEOS would,in a similar vein, benefit from a more complex model that accounts for the thermal properties of various components such as solar panels, radiators and others. 
Concerning radiation, total ionization dose effects are not currently implemented in PASEOS, but they will become relevant for simulations in the time scales of years \cite{furano2020towards}, especially for commercial off-the-shelf devices \cite{JetsonNanoRadhard, furano2020towards, lentaris2018high}.

Another direction of improvements lies in expanding the the capabilities of actors. Ground-based actors are currently stationary and only supported for scenarios on Earth. Further, space-based actors cannot perform manoeuvres (although one can manually change the orbit). In the future, these capabilities could support more complex operational scenarios and activities. 

Overall, there is a virtually endless range of potential improvements in fidelity and modelled aspects. It will require careful analysis of which aspects are critical to enable realistic constraint consideration and which ones can be simplified to the degree as is currently the case in PASEOS. But, to even enable these comparisons, one needs to start with a baseline first, which is what we are providing here. 

\section{Conclusion}
PASEOS is a software package that enables holistic modelling of the onboard environment accounting for, e.g., power, thermal, radiation, and dynamics.
The generality of PASEOS makes it a tool to study a plethora of operational scenarios, hardware configurations, or to be used in conjunction with other simulation tools. 
In particular, PASEOS is well suited to study constellations in space for emerging operational scenarios, such as edge computing, edge and decentralized learning, and artificial intelligence in space \cite{sanchez2022edge}.

Overall, we have demonstrated that PASEOS provides the means to model a variety of constraints that spacecraft and their operators experience in orbit. 
Thus, we can explore the feasibility of onboard activities with a greater rigor before launch and/or form an understanding of how already operational assets may be repurposed or perform in the future. 
Given PASEOS's asynchronous and versatile setup, a broad range of scenarios ranging from one to multiple spacecraft (including ground-based actors) are possible. Both real-time at-the-edge execution and long-term simulation on a computing cluster or similar infrastructure are supported. The specifics of the modelled quantities can be adjusted to fit the particular scenario. 

However, there are of course intrinsic limitations to this process. At the moment, the models inside PASEOS exhibit comparatively lower fidelity than simulators for dedicated topics (e.g. \textit{ns-3} or \textit{OMNeT++})  to enable rapid background execution. PASEOS's modular nature does provide a natural surface for extension and wrapping of more complex physical models. In the future, we will conduct application-specific studies, explore more complex models, and demonstrate an operational scenario using PASEOS on multiple edge devices in parallel to solve a real-world task. 

\section*{Acknowledgment}
The authors would like to thank Unibap AB for providing the iX-10 100 device that was used for our experiments. 

\bibliographystyle{unsrt}
\bibliography{references}



\EOD

\end{document}